\documentclass[aps,prc,showpacs,floatfix,preprint]{revtex4}
\usepackage{amsmath,bm}
\usepackage{graphicx}
\usepackage{epsfig}
\begin{document}

\title{Hadron production from resonance decay in relativistic collisions}
\author{Subrata Pal and Pawel Danielewicz}
\affiliation{National Superconducting Cyclotron Laboratory
and Department of Physics and Astronomy, Michigan State University, East
Lansing, Michigan 48824}

\begin{abstract}
A statistical model for decay and formation of heavy hadronic resonances
is formulated. The resonance properties become increasingly uncertain with
increasing resonance mass. Drawing on analogy with the situation in low-energy nuclear
physics, we employ the Weisskopf approach to the resonance processes.
In the large-mass limit, the density of resonance states in mass is governed by
a universal Hagedorn-like temperature $T_H$. As resonances decay, progressively
more and more numerous lighter states get populated. For $T_H \simeq 170\, \mbox{MeV}$,
the model describes data for the hadron yield ratios at the RHIC and SPS
energies under the extreme assumption of a single heavy resonance giving rise to
measured yields.
\end{abstract}
\maketitle

Studies of ultrarelativistic nuclear reactions aim at learning on
properties of highly-excited strongly-interacting matter and, in particular,
on the predicted transition to quark-gluon plasma (QGP) \cite{qm2004,gyulassy}.
Yields from those reactions can be described in terms of grand-canonical
thermodynamic models \cite{braun,redlich,becattini,braun1}, at temperatures close to those
anticipated for the transition \cite{karsch}. Similar temperatures are utilized
in the microcanonical model for electron-positron annihilation into hadrons
\cite{ep}. Large fractions of final light-particles within those thermodynamic equilibrium
models result from secondary decays of heavy resonances whose features are, generally,
less and less known the heavier the resonances. The lack of knowledge has forced the use of cut-offs
on the primary resonance mass in equilibrium models \cite{braun,redlich,becattini,braun1};
analogous cut-offs have been employed in transport models
\cite{urqmd,hsd,ampt}. In this paper, we explore the
possibility of an apparent equilibrium in the reactions stemming from
sequential binary decays of heavy resonances, with the characteristic
temperature describing the resonance mass spectrum, rather than being proposed {\em ad hoc}.

The situation of deteriorating knowledge of resonance properties, with increasing
resonance energy and resonance density in energy, reminisces that of resonances
in low-energy nuclear physics. There, statistical descriptions, in terms of the
Weisskopf compound-nucleus model and the Hauser-Feschbach theory, have been highly successful
\cite{statmod}. Underlying the statistical descriptions is the density of resonance
states in energy $\rho(m,q)$, where $q$ represents discrete quantum numbers of the
resonances. For the hadronic states, a universal temperature in terms of the density of states,
$T_H^{-1} = \partial \log{\rho}/ \partial m$, has been originally considered by Hagedorn
\cite{hagedorn}, in part due to an evidence for a rapid, nearly exponential, increase
in the number of resonances with energy; see also \cite{broniowski}. To reach a
conclusion on the basis of spectrum, however, the counting of resonances should be
carried out at constant values of the quantum numbers~$q$, particularly in the region
of opening thresholds for different~$q$. This has not been considered in Refs.\
\cite{hagedorn,broniowski}. Nonetheless, in the limit of large $m$ at fixed~$q/m$, the universal
temperature $T_H$, independent of $m$, but possibly dependent on $q/m$, may be expected
for the resonances due to the lack of any scales that could govern the $T_H$-dependence
on $m$.  While our model for $\rho$ will be fairly schematic, similar to the models
employed in the literature \cite{hagedorn,broniowski}, there are generally important
questions regarding strong interactions that can be suitably asked in terms of $\rho$,
concerning e.g., besides the $T_H(q/m)$ dependence, the emergence of a surface tension
in the thermodynamic limit.
In the context of the phase transition, the temperature $T_H$ may be considered as the
temperature for a metastable equilibrium of quark-gluon drops with vacuum, and, as such,
slightly lower temperature than the critical~$T_c$.

For resonances described by the continuum density of states $\rho$, we consider
the processes of binary breakup and inverse fusion, constrained by
detailed balance, geometry and by the conservation of baryon number $B$, strangeness $S$,
isospin $I$ and of isospin projection~$I_z$. From the two versions of our model,
with and without a strict conservation of angular momentum~$(J,J_z)$,
we discuss the simpler Weisskopf version, in this first model presentation.
The lighter particles (comprising of 55 baryonic and 34 mesonic states) are treated
explicitly as discrete states in the model.  Our model allows to explore various aspects of the
system evolution in ultrarelativistic collisions, including formation of hadrons
with extreme strangeness and isospin, as well as chemical and kinetic freeze-out.

We first discuss details of the density of states.
As a threshold mass, separating the discrete states from those
in the statistical continuum, we take $m_{\rm th}^c= 2 \, \mbox{GeV}$. We assume that
at low excitation energies the density of states is similar at different $q$ for
the same excitation energy above the spectrum bottom $m_g(q)$.
Following quark considerations, we adopt
\begin{equation}
\label{gsmas}
m_g(q) = a_Q \, {\rm max}\left(|3B+S|,2I\right) + a_S \, |S| \, ,
\end{equation}
where $a_Q=0.387 \, \mbox{GeV}$ and $a_S=0.459 \, \mbox{GeV}$.  The coefficient magnitudes have
been adjusted requiring that the mass $m_g(q)$ from Eq. (\ref{gsmas})
{\em exceeds} the lowest known masses for different~$q$; regarding the practicality
of sequential decays, we prefer to overestimate rather than to underestimate the reference
position of spectrum bottom for the continuum, and to overestimate its rise with~$q$,
to preclude the emergence of any unphysical stable states towards the continuum bottom.
At high masses $m$ the influence of the ground state mass on $\rho$ should decrease.
We eventually arrive at the following density of states employed in our calculations:
\begin{equation} \label{dos}
\rho(m,q) = A \, \frac{\exp\!\left[\left\{m-m_g \, f(m-m_g)\right\}/T_H\right]}
{\left[\left\{m-m_g \, f(m-m_g)\right\}^2 + m_r^2\right]^\alpha} \, .
\end{equation}
The role of the factor $f$ is to suppress the effect of $m_g$ at high $m$ and we use
\begin{equation} \label{scaling}
f(m-m_g) = \frac{1}{1+\left[(m-m_g)/m_{\rm th}^c\right]^n} \, .
\end{equation}
with $n=1$. The prefactor of the exponential in Eq.~(\ref{dos}), with $m_r=0.5 \, \mbox{GeV}$, acts to
enhance asymmetric (rather than symmetric) breakups for continuum hadrons, leaving the
issue of surface tension in the thermodynamic limit open. The normalizing factor $A$
and the power $\alpha$ are adjusted, for different assumed values of Hagedorn temperature
$T_H$, by comparing the low-$m$ cumulant spectra from measurements and from the continuum
representation~\cite{broniowski}:
\begin{eqnarray}\label{cumulant}
N_{\rm exp} &=& \sum_i (2J+1) \, \Theta(m-m_i) \, , \nonumber\\
N_{\rm the} &=& \sum_q \int^m {\rm d}m' \, \rho(m',q) \, .
\end{eqnarray}
For $T_H = 170$ MeV, as an example, we obtain the prefactor power of $\alpha=2.82$.
The normalization factor $A$ drops out from probabilities for the most common processes
involving continuum hadrons, with one continuum and one discrete hadron
either in the initial or final state.

We now turn to cross sections and decay rates.
The cross section for the formation of a~resonance $q$ in the interaction
of hadrons 1 and~2, can be, on one hand, represented as~\cite{danielewicz}:
\begin{eqnarray}\label{exsec}
\sigma(q_1+q_2 \to q) &=& \frac{1}{v_{12}} \, \frac{m_1}{e_1({\bf p}_1)} \,
\frac{m_2}{e_2({\bf p}_2)} \int \frac{{\rm d} {\bf p}}{(2\pi)^3}
\int \frac{{\rm d} e}{2\pi} \, 2 \pi \, \rho (m,q) \,
|{\cal M}_{q_1+q_2 \to q}|^2 \nonumber\\
&~~& \times (2\pi)^3 \, \delta\left({\bf p}_1 + {\bf p}_2 - {\bf p} \right)
2\pi \, \delta \left( e_1({\bf p}_1) + e_2({\bf p}_2) - e \right) \nonumber\\
&=& \frac{2 \pi \, m_1 \, m_2}{m \, p^\star (m_1,m_2) } \, \rho (m,q) \,
|{\cal M}_{q_1+q_2\to q}|^2 \, .
\end{eqnarray}
Here, $v_{12}$ is the relative velocity, $e$'s are the single-particle energies, and
$|{\cal M}|^2$ is the matrix element squared for the fusion, which is averaged over
initial {\em and} final spin directions. The factor of $(2J + 1)$, associated
with the last averaging, has been absorbed into $\rho$.  The c.m.\ momentum
in Eq. (\ref{exsec}) is
\begin{equation} \label{momentum}
p^\star (m_a,m_b) = \frac{1}{2m} \left[ \left(m^2-(m_a+m_b)^2\right)
\left(m^2-(m_a-m_b)^2\right) \right]^{1/2} \, .
\end{equation}
For the final state in continuum,
following geometric considerations, the cross
section for fusion, on the other hand, is
\begin{eqnarray}\label{gxsec}
\sigma(q_1+q_2 \to q) =  \langle I_1 \, I_{z_1} \, I_2 \, I_{z_2} || I \, I_{z} \rangle
\, \pi \, R^2 \,  ,
\end{eqnarray}
where $\langle \cdot || \cdot \rangle$ represents the isospin Clebsch-Gordan coefficient.
The cross section radius is taken as that of the fused entity, $R \approx r_0 \, (m/m_d)^{1/3}$, with
$r_0 = 1 \, \mbox{fm}$ as a characteristic radius for a $m_d = 1 \, \mbox{GeV}$ hadron.
Equations (\ref{exsec}) and (\ref{gxsec}) allow to extract
the square of the transition matrix element needed for computation of the partial
resonance-width.

The partial width for decay into $q_1$ and $q_2$ can be generally represented as
\begin{eqnarray}\label{width}
\Gamma (q \to q_1 + q_2) &=& \int \frac{{\rm d} {\bf p} }{ (2 \pi)^3 }
\int {\rm d} m_1' \, \frac{m_1' \, \rho (m_1',q_1)}{e_1({\bf p})}
\int {\rm d} m_2' \, \frac{m_2' \, \rho (m_2',q_2)}{e_2({\bf p})} \nonumber\\
&~~&\times  |{\cal M}_{q \to q_1 + q_2}|^2 \, 2 \pi \,
\delta\left( e_1({\bf p}) + e_2({\bf p}) - m \right) \, .
\end{eqnarray}
As before, the factors of $(2J_j+1)$ are absorbed into $\rho_j$.
A resonance in the continuum can undergo three types of binary decay: where
(i) both daughters are particles with well established properties within the discrete
spectrum below $m_{\rm th}^c= 2 \, \mbox{GeV}$, (ii) one of the daughters belongs
to the discrete spectrum and the other to the continuum and, finally, where (iii) both
daughter resonances belong to the continuum. In the case (i),
the state densities are $\rho_j(m_j')=(2J_j+1) \, \delta(m_j' - m_j)$. With the detailed
balance relation, $|{\cal M}_{q \to q_1 + q_2}|^2 = |{\cal M}_{q_1+q_2\to q}|^2$, we
then get
\begin{eqnarray}\label{ddwidth}
\Gamma^{(i)} (q \to q_1 + q_2) &=& \frac{m_1 \, m_2 \, p^\star(m_1,m_2)}{\pi \, m} \,
(2J_1+1) \, (2J_2+1) \, |{\cal M}_{q \to q_1 \, q_2}|^2 \nonumber\\
&=&
\langle I_1 \, I_{z_1} \, I_2 \, I_{z_2} || I \, I_{z} \rangle \,
\frac{(2J_1+1) \, (2J_2+1) \, p^{\star 2}(m_1,m_2) \, R^2}{ 2 \pi \, \rho(m,q)} \, .
\end{eqnarray}

In analyzing the case (ii), let the particle characterized by $q_1$ belong to the
discrete spectrum and that characterized by $q_2$ to the continuum
spectrum. From (\ref{width}), we then find
\begin{eqnarray}\label{dc1width}
\Gamma^{(ii)} (q \to q_1 + q_2) &=&
\langle I_1 \, I_{z_1} \, I_2 \, I_{z_2} || I \, I_{z} \rangle \,
\frac{(2J_1+1)\, m \,R^2 \, } {2 \pi \, \rho(m,q)} \nonumber \\
&~~&\times \int_0^{p^*(m_1,m_{c_2})} {\rm d}p \,
\frac{p^{ 3} \, \rho(\sqrt{m^2+m_1^2-2m \, e_1},q_2) }{e_1 \,
\sqrt{m^2+m_1^2-2m \, e_1}} \, ,
\end{eqnarray}
where $m_{c_2}= \mbox{max}(m_{\rm th}^c,m_g(q_2))$.  For $e_1$ small compared to $m$,
it is of advantage to represent the subintegral density of states as an exponential
of the density logarithm and to expand the logarithm in $(e_1-m_1)$. The
integration over relative momentum can be thereafter carried out explicitly,
yielding
\begin{equation} \label{dcwidth}
\Gamma^{(ii)} (q \to q_1 + q_2) =
\langle I_1 \, I_{z_1} \, I_2 \, I_{z_2} || I \, I_{z} \rangle \,
\frac{(2J_1 + 1) \, m R^2 \, T_{2}^2 \, (T_{2}+m_1)  }{\pi \, m_2}
\, \frac{ \rho(m_2,q_2)}{\rho(m,q)} \, .
\end{equation}
Here, $m_2=m-m_1$ and the temperature is defined as
\begin{eqnarray}\label{temper}
\frac{1}{T_2} & = & \frac{m}{m_2} \, \frac{\partial \, \log{\rho(m_2, q_2)}}
{\partial m_2} \nonumber \\
& \approx & \frac{m}{m_2 \, T_H} \, \left\lbrace
1 + \frac{ n \, (m^c_{\rm th})^n \, \left[ m_2 - m_g(q_2) \right]^{n-1} }
{(m^c_{\rm th})^n + \left[ m_2 - m_g(q_2) \right]^{n}}
\right\rbrace
 \, .
\end{eqnarray}
For $(m_1,q_1)$ small compared to $(m,q)$, the obvious further possibility is
the expansion of $\log{ \rho(m_2,q_2)}$ in $(m_1,q_1)$, with an emergence of
the chemical potentials conjugate to $q_2$.

In the case (iii), of both daughters in the continuum with large masses
$m_i > m_{th}^c = 2 \, \mbox{GeV}$, the nonrelativistic limit in Eq.~(\ref{width}) is justified.
On employing Eqs.~(\ref{exsec}) and (\ref{gxsec}), the result for the partial width is
\begin{eqnarray}\label{ccwidth}
\Gamma^{(iii)} (q \to q_1 + q_2) & = &
\langle I_1 \, I_{z_1} \, I_2 \, I_{z_2} || I \, I_{z} \rangle
\, \frac{m \, R^2}{8 \pi \, \rho(m,q)} \, \int_0^{m-m_{c_1}-m_{c_2}} {\rm d} \epsilon \,
\epsilon \, \overline{\rho}(m - \epsilon, q_1, q_2) \nonumber \\[.5ex]
& = &
\langle I_1 \, I_{z_1} \, I_2 \, I_{z_2} || I \, I_{z} \rangle
\, \frac{m \, R^2 \, T_{12}^2}{8 \pi} \, \frac{\overline{\rho}(m, q_1, q_2)}{ \rho(m,q)}
\, ,
\end{eqnarray}
where
\begin{equation}
\overline{\rho}(m, q_1, q_2) = \int_{2m_{c_1}-m}^{m-2m_{c_2}} {\rm d}u \,
\rho \left({(m+u)}/{2},q_1 \right) \, \rho \left( {(m-u)}/{2},q_2 \right)
\, \left(1 - {u^2}/{m^2} \right) \, .
\end{equation}
To obtain the last expression in Eq. (\ref{ccwidth}), we have expanded the logarithm
of subintegral density, with the temperature representing
$1/T_{12}= \partial [\log\overline{\rho}(m, q_1, q_2)]\big/ \partial m$.

The total decay width of an $(m,q)$ resonance is finally
\begin{equation}\label{twidth}
\Gamma(m,q) = \sum_{q_1 \, q_2} \Gamma^{(i)}(q \to q_1 + q_2)
+ \sum_{q_1 \, q_2} \Gamma^{(ii)}(q \to q_1 + q_2)
+ \sum_{q_1 \, q_2} \Gamma^{(iii)}(q \to q_1 + q_2) \, .
\end{equation}
A moving resonance will live an average time of $\langle \tau \rangle = \gamma
/ \Gamma$, where $\gamma$ is the resonance Lorentz factor.

The formulas above provide the basis for our Monte-Carlo simulations of the resonance decay
sequences in heavy ion collisions. A resonance follows an exponential decay law corresponding to
$\langle \tau \rangle$. The product properties are selected according to the
decay branching ratios. Since the parent angular momentum is not tracked in the Weisskopf approach,
the angular distribution of products is taken as isotropic.  During the evolution two resonances can
fuse with each other, according to the cross section of Eq.\ (\ref{gxsec}), if the final state is
in continuum.  If, on the other hand, the state is discrete, the cross section acquires the standard
form \cite{danielewicz}, from Eqs.\ (\ref{exsec}) and (\ref{ddwidth}),
\begin{equation}\label{ddreco}
\sigma(q_1+q_2 \to q) = \frac{2\pi^2 \,\rho(m,q) \, \Gamma (q\to q_1+q_2) }
{(2J_1+1) \, (2J_2+1) \, p^{\star 2}(m_1,m_2)}
\,
,
\end{equation}
where the width for the spectral function is considered:
\begin{equation}
\rho(m,q) =  \frac{(2J+1) \, 2m_q^2 \, \Gamma}
{\pi  \left[(m^2 - m_q^2)^2 + m_q^2 \, \Gamma^2 \right]}
\approx \frac{(2J+1) \, \Gamma}{2\pi  \left[(m - m_q)^2 + \Gamma^2/4\right]}
\, .
\end{equation}
Besides the decay and fusion processes, related by detailed balance, a provisional
constant cross section $\sigma_{\rm el} = 5$~mb has been assumed for all collisions.

In our model, we simulate, in particular, the features of the final
state of central Au+Au collisions at $\sqrt s = 130A$ GeV. Following the presumption
that the decay sequences will tend to erase fine details of the initial state, we push
the characteristics of the initial state to an extreme, allowing for a single heavy resonance
to populate a given rapidity region. In the end, when examining the transverse
momentum spectra, we find that the resonance decay and reformation alone generates
insufficient transverse collective energy, indicating that the early resonances need
to be affected by a collective motion generated prior to the resonance
stage. This finding is consistent with those in other works \cite{braun,heinz}.

Within the single resonance scenario, the local final state reflects the initial
quantum numbers of a resonance characterized by $(m_0, q_0)$, where
$q_0 \equiv (B_0,S_0,I_0,I_{z \, 0})$.  After the value of the Hagedorn
temperature $T_H$ is set, the relative yields of particles in the final
state are, in practice, sensitive only to the relative values of the quantum numbers of
the initial resonance. We normally impose strangeness neutrality, so that the starting
value is $S_0=0$. The magnitude of $B_0$, for a given $m_0$, can be adjusted by using the
final-state antiproton-to-proton or antiproton-to-pion ratios.
The starting isospin, for a given $B_0$, can be adjusted by using the
isospin of original nuclei. However, we find that the initial isospin has only a~marginal impact
on the isospin of individual final particles. This may be attributed to the
cumulative effect of isospin fluctuations when many particles, compared to~$I$, are
produced. For specific initial $(m_0, q_0)$ values, we repeat numerous times
the Monte-Carlo simulations of the decay chain and recombination, and the results
presented here are an average of about~$10^4$ generated event sequences.

\begin{figure}[th]
\centerline{\epsfig{file=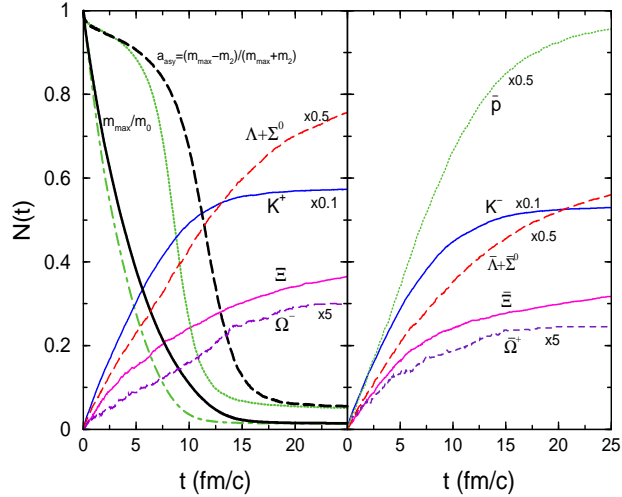,width=3.2in,height=2.6in,angle=0}}
\caption{
Time evolution of average characteristics for a system starting at $m_0=100$~GeV,
$B_0=3$ and $S_0=0$, under the assumption of $T_H=170$~MeV.  Displayed in the left
panel is the ratio of the maximal resonance mass to the initial resonance mass, $m_{\rm max}/m_0$,
as well as the mass asymmetry, $a_2=(m_{\rm max}-m_2)/(m_{\rm max}+m_2)$, where $m_2$
is the mass of the second heaviest resonance. These results are shown both for the standard
system evolution, then represented by the monotonically dropping solid and dashed lines,
and for the evolution with suppressed fusion reactions, then represented by the monotonically
dropping dash-dotted and dotted lines, respectively. In addition, the left
and right panel display the evolution of particle and antiparticle abundances, respectively,
for the standard evolution, in terms of rising lines.
}
\label{tyield}
\end{figure}

Figure \ref{tyield} illustrates the average features of an exemplary local system
that starts out as a resonance characterized by $m_0=100$~GeV and $B_0=3$.
The left panel shows the ratio of the average maximal resonance mass $m_{\rm max}$
to the initial mass $m_0$, as a function of time, as well as the mass asymmetry,
the mass difference between the heaviest resonance and the next heaviest,
divided by the sum of their masses, $a_2=(m_{\rm max}-m_2)/(m_{\rm max}+m_2)$.
Persistence of the large asymmetry with time indicates that the heavy resonance decays
primarily through light-hadron emission.

In addition, Fig. \ref{tyield} shows, as a function of time, the abundances of particles
(left panel) and antiparticles (right panel). The antibaryon abundances freeze out
noticeably earlier than the baryon abundances, in spite of a low initial
baryon number relative to the initial mass. The higher the strangeness, the later
the abundance saturates. This is likely due to the fact that the effects of strangeness
fluctuation need to accumulate with time; the situation would change if we assumed
strangeness fluctuations right for the initial resonance conditions.

The calculations have been repeated while suppressing the back resonance fusion
reactions. As expected, in this case the maximal mass and asymmetry decrease faster
with time, see the left panel of Fig. \ref{tyield}. The abundances (not shown)
grow faster and saturate earlier for the modified evolution.

\label{ratios}
\begin{table}
\caption{Particle yield ratios from the resonance decay model compared to the RHIC
Au+Au central-collision data at $\sqrt{s}=130 \, A$~GeV. In the model, the starting
baryon number of $B_0=3$ is assumed, and the mass of either $m_0=100$~GeV or $m_0=95$~GeV
with a corresponding resonance mass spectrum temperature of either $T_H=170$~MeV or
$T_H=175$~MeV is used. Results obtained when suppressing fusion reactions, for
$T_H=170$~MeV, are provided in parenthesis.
}
\begin{tabular}{l l l l l l l}\hline
Yield-Ratio          & \multicolumn{2}{c}{Decay-Model Results} & \multicolumn{1}{c}{Experimental}
& Collaboration & Ref. \\
               & $T_H=170$ MeV & $175$ MeV & \multicolumn{1}{c}{Data} & &  \\ \hline

$\overline{p}/\pi^{-}$  & 0.06 (0.05) & 0.07& 0.07 $\pm$ 0.01   & STAR & \cite{olga} \\

$\pi^{-}/\pi^{+}$       & 1.01 (1.01) & 1.01 & 1.00 $\pm$ 0.02 & PHOBOS & \cite{ratioph} \\[-3mm]
                        &      &  & 0.95 $\pm$ 0.06 & BRAHMS & \cite{bearden}  \\

$K^+/K^-$               & 1.073 (1.078) & 1.074 & 1.092 $\pm$ 0.023   & STAR & \cite{kpkmst} \\ [-3mm]
                        &       &  & 1.28 $\pm$ 0.13 & PHENIX & \cite{zajc}     \\ [-3mm]
                        &       &  & 1.09 $\pm$ 0.09 & PHOBOS & \cite{ratioph} \\ [-3mm]
                        &       &  & 1.12 $\pm$ 0.07 & BRAHMS & \cite{bearden}  \\

$K^-/\pi^-$         & 0.175 (0.185) & 0.175 & 0.146 $\pm$ 0.024 & STAR & \cite{kpist} \vspace{0.2cm} \\


$\overline{p}/p$        & 0.65 (0.63) & 0.66 & 0.65 $\pm$ 0.07   & STAR & \cite{pbpst} \\ [-3mm]
                        &       &  & 0.64 $\pm$ 0.07 & PHENIX & \cite{zajc} \\ [-3mm]
                        &       &  & 0.60 $\pm$ 0.07 & PHOBOS & \cite{ratioph} \\ [-3mm]
                        &       &  & 0.64 $\pm$ 0.07 & BRAHMS & \cite{pbpbh}  \\

$\overline{\Lambda}/\Lambda$  & 0.72 (0.69) & 0.73 & 0.71 $\pm$ 0.04  & STAR & \cite{kpkmst} \\[-3mm]
                              &      &         & 0.75 $\pm$ 0.19  & PHENIX & \cite{lblapx} \\
$\overline{\Xi}^+/\Xi^-$ & 0.81 (0.73) & 0.80 & 0.83$\pm$ 0.06  & STAR & \cite{kpkmst} \\
$\overline{\Omega}^+/\Omega^-$  & 0.82 (0.76) & 0.82
                                &  0.95 $\pm$ 0.15 & STAR & \cite{kpkmst} \vspace{0.2cm} \\ %

$\Lambda/h^-$     & 0.054 (0.048) & 0.060 & 0.054 $\pm$ 0.015   & STAR  &  \cite{lblast} \\
$\overline\Lambda/h^-$     & 0.040 (0.034) & 0.046 & 0.040  $\pm$ 0.015  & STAR  &  \cite{lblast} \\
$\Lambda/p$   & 0.57 (0.58) & 0.59 & 0.89 $\pm$ 0.22  & PHENIX  & \cite{lblapx}  \\
$\overline\Lambda/\overline p$   & 0.63 (0.63) & 0.66 & 0.95 $\pm$ 0.24  & PHENIX &  \cite{lblapx} \vspace{0.2cm} \\
$\Xi^-/h^-$            & [6.0 (5.0)]$\times 10^{-3}$
                       & 6.9$\times 10^{-3}$
                       & [7.9$\pm$1.1]$\times 10^{-3}$   & STAR  &  \cite{mstrst} \\
$\overline{\Xi}^+/h^-$  & [4.8 (3.8)]$\times 10^{-3}$
                        & 5.6$\times 10^{-3}$
                        & [6.6$\pm$0.8]$\times 10^{-3}$   & STAR  & \cite{mstrst} \\
$\Xi^-/\pi^-$        & [7.5 (6.7)]$\times 10^{-3}$
                     & 8.9$\times 10^{-3}$
                     & [8.8$\pm$0.4]$\times 10^{-3}$ & STAR  & \cite{xipist} \\
$\Xi^-/\Lambda$                     & 0.108 (0.109) & 0.115 & 0.193 $\pm$ 0.032  & STAR  & \cite{mstrst} \\
$\overline{\Xi}^+/\overline\Lambda$ & 0.118 (0.114) & 0.122 & 0.219 $\pm$ 0.037  & STAR  & \cite{mstrst} \vspace{0.2cm} \\
$\Omega/h^-$            & [2.7 (2.4)]$\times 10^{-3}$
                        & 3.2$\times 10^{-3}$
                        & [2.2$\pm$0.6]$\times 10^{-3}$   & STAR  &  \cite{mstrst} \\ \hline


\end{tabular}
\label{table:ratios}
\end{table}

Table \ref{table:ratios} compares the yield ratios from the resonance-decay model,
within the region of optimal model-parameter values, with the available central
Au+Au collision data from RHIC at $\sqrt{s}=130A$ GeV \cite{feed}. The temperature of the
resonance mass spectrum, $T_H$, and the initial baryon-number-to-mass ratio, $B_0/m_0$,
can be adjusted by simultaneously considering the ratios of $p/\pi^+$ and $\bar p/p$.
An optimal agreement between the model and the central RHIC data is obtained for
$T_H \approx 170$ MeV and $B_0/m_0 \approx 0.030$~GeV$^{-1}$. The model results tend
to be only weakly sensitive to $m_0$:  with an increase in $m_0$ a slight increase in
$B_0/m_0$ is favored, that can be traced to the factor $f(m-m_g)$ in the density of states.

The overall agreement between data and the decay-model calculations in Table
\ref{table:ratios} is quite remarkable, given that only two parameters,
$T_H$ and $B_0/m_0$, are adjusted. One should note
that the optimal Hagedorn temperature of $T_H = 170$~MeV is close to the
critical temperature of $T_c\sim 170$ MeV for a transition to the color-deconfined
QGP phase obtained in the lattice QCD calculations at zero net baryon density
\cite{karsch}, and it is also similar to the chemical freeze-out
temperature $T_{\rm ch} = 174$ MeV extracted from the analysis of RHIC
data within a grand-canonical model \cite{braun}.

At the general level, the calculations are quite good in reproducing yield ratios
involving strange particles.  A more detailed examination, however, reveals some potential deficit
of multistrange baryons and antibaryons. Possible reasons for the deficiency, to be
investigated in the future, include: the possible role played by the
non-resonant strangeness-exchange processes  \cite{pal}
and by the multiparticle processes \cite{carsten} and, further,
the possible sensitivity of strangeness production to an early system dynamics
\cite{koch} and, specifically, to strangeness fluctuations for early resonances
and/or to details in flavor-dependence of state density.

Besides the yield ratios in the standard model evolution, the corresponding ratios
from the evolution with suppressed fusion processes are given in parenthesis in Table
\ref{table:ratios} for $T_H = 170$~MeV. Though the suppression of fusion alters particle
abundances early on in the system development, the final yields turn out to be rather
similar, quite uniformly across the particle species. Only a careful examination
reveals that the fusion suppression enhances slightly the production of pions and
other light mesons and reduces slightly the production of heavier strange particles.

Within a moderate range, there is no strong preference for a particular Hagedorn
temperature.  As Table \ref{table:ratios} shows, similar yield ratios are obtained
for $T_H = 175$~MeV as for $T_H = 170$~MeV, if, in the context of the RHIC data, a slightly
reduced initial mass of $m_0=95$~GeV is assumed.

\begin{figure}[ht]
\centerline{\epsfig{file=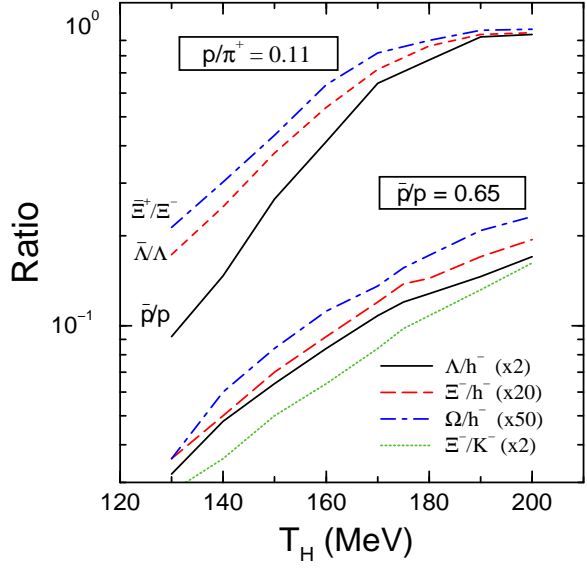,width=3.0in,height=3.0in,angle=0}}
\caption{Dependence of particle yield ratios on Hagedorn temperature for a
fireball characterized by the $B_0/m_0$ ratio adjusted to reproduce either the
yield
ratio of $p/\pi^+ =0.11$ (upper set of lines) or $\overline p/p = 0.65$ (lower set of lines).
The ratios chosen for the adjustment
represent the
central Au+Au data at
$\sqrt s = 130A$ GeV.}
\label{tempratio}
\end{figure}

In Fig.~\ref{tempratio}, we show the particle yield ratios that turn out to
be particularly sensitive to the Hagedorn temperature under a given constraint.
In one case, while varying the temperature, we adjust the initial fireball's ratio
$B_0/m_0$ to reproduce the ratio $\bar p/p \approx 0.65$ for the RHIC
data (lower set of lines). In that case, the baryon-to-antibaryon ratios, $\bar B/B$,
remain rather stable with $T_H$ variation; the strongest variations are observed
for the ratio of strange baryons to the negatively charged hadrons or to the negative mesons.
On the other hand, if we adjust the ratio $B_0/m_0$ to reproduce the ratio
$p/\pi^+ \approx 0.11$ (upper set of lines), strong variations are observed for
the $\bar B/B$ ratios.  No matter what fitting strategy is followed, a reasonable
agreement with the data is obtained within the Hagedorn temperature range
of $T_H = 165-175\, \mbox{MeV}$.

\begin{figure}[ht]
\centerline{\epsfig{file=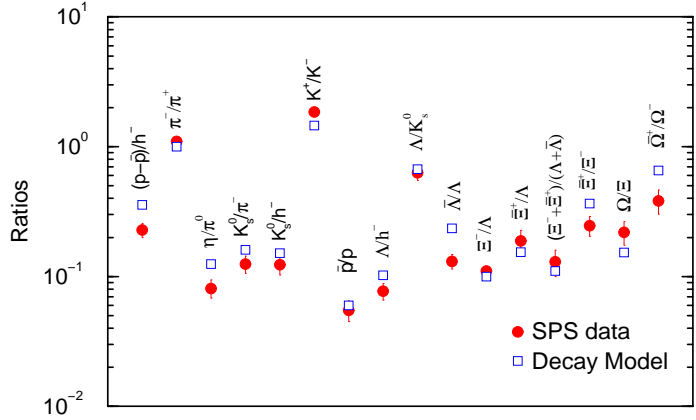,width=3.6in,height=2.1in,angle=0}}
\caption{Particle yield ratios from the resonance decay model compared to the
SPS Pb+Pb central-collision data at the laboratory energy of
$158A$~GeV.  In the model, at $T_H=170$ MeV, the starting baryon number of $B_0=26$ was
assumed, in combination with the mass $m_0=100$~GeV and strangeness $S_0=0$.}
\label{spsratio}
\end{figure}

We next confront our resonance decay model with the SPS abundance data from the
central Pb+Pb collisions at the laboratory energy of $158A$~GeV.
As illustrated in Fig. \ref{spsratio}, an optimal agreement with the data is obtained
for $m_0=100$~GeV when assuming (at $T_H=170$~MeV) a starting baryon number of $B_0=26$.
While the general
agreement is rather good, we note that the calculated ${\overline\Omega}^+/\Omega^-$
ratio is about $60\%$ larger than the data.  It is likely that the assumption of a larger
number of lighter initial resonances would improve the agreement; in the thermal model
the discrepancy is tauted as strangeness undersaturation~\cite{becattini}.

Measured kinematic spectra of particles from central heavy ion collisions exhibit the effects
of collective expansion. As may be expected, with suppressed fusion processes, our model 
produces kinematic spectra characterized by slope temperatures close to 
$T_H$ (see also Ref.~\cite{moretto}). In the standard version of the model, the 
sequences of decay and fusion generate some collective motion, but not enough to explain 
the transverse RHIC spectra. For pions e.g.\ the slope temperature raises by
4\% compared to the version without fusion.  Moderate increases in the elastic cross section,
such as to an overall 10~mb, raise kinematic temperatures further, 6\% for pions, but not well
enough to approach data.  As a consequence, it is necessary to assume the presence of some
collective motion early on in system evolution, leading to resonances that exhibit
space-momentum correlations. It might be that the dynamics, beyond statistics, needs to be
involved in the predominant decays, involving the interior degrees of freedom of
resonances. The first resonances might also emerge at finite transverse velocities.
In either case, degrees of freedom beyond resonances would be involved.

\begin{figure}[ht]
\centerline{\epsfig{file=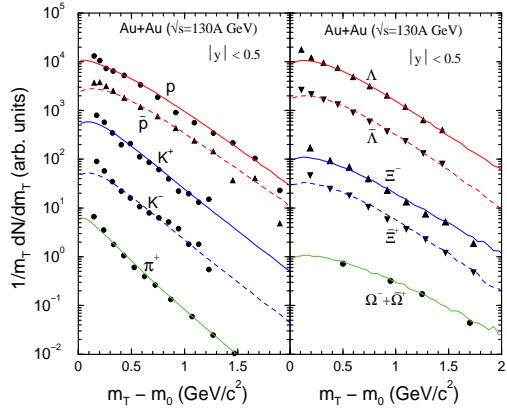,width=2.6in,height=2.1in,angle=0}}
\caption{Transverse mass spectra of midrapidity hadrons at RHIC. Solid symbols are
the data from~PHENIX~\protect\cite{sppx} (left panel) and STAR
\protect\cite{lblast,mstrst} (right panel) collaborations from central Au+Au
collisions at the energy of $\sqrt s = 130A$~GeV. The lines represent
the model spectrum from convoluting the spectrum of the resonance decay
model, at $T_H=170$~MeV, with a uniform transverse collective velocity
distribution characterized by $\langle \beta_t\rangle = 0.27$.}
\label{mtspectra}
\end{figure}

\begin{figure}[ht]
\centerline{\epsfig{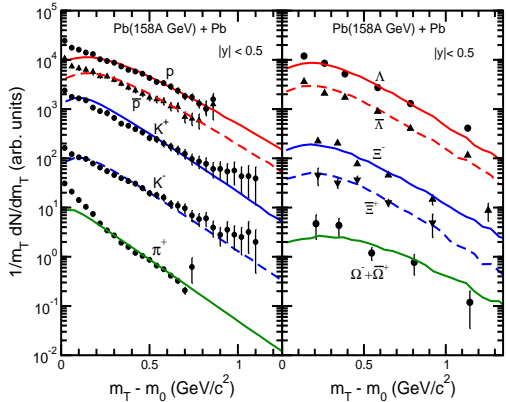}}
\caption{Transverse mass spectra of midrapidity hadrons at SPS.  Solid symbols are
the data from~NA44~\protect\cite{na44}(left panel) and WA97 \protect\cite{wa97}
collaborations from central Pb+Pb collisions at the energy of 158 AGeV.
The lines represent the model spectrum from convoluting the spectrum of 
the resonance decay model, at $T_H=170$~MeV, with a uniform transverse 
collective velocity distribution characterized by $\langle \beta_t\rangle = 0.14$.}
\label{spsmtspectra}
\end{figure}

The transverse mass spectra displayed in Fig.~\ref{mtspectra} are obtained by
folding a common collective velocity field with the spectra from our decay model.
Specifically, we assume a uniform transverse velocity distribution,
$d^2N/d\beta_t^2 = \Theta(\beta_t-\beta_{\rm max})$. The spectra for all the hadrons
can be best described, at $T_H=170$~MeV, with a uniform velocity field of
$\beta_{\rm max} \approx 0.40$, corresponding to an average flow velocity of
$\langle \beta_t\rangle = 2 \, \beta_{\rm max}/3 \approx 0.27$.
Notably, much less early flow, as characterized by
$\langle \beta_t\rangle \approx 0.14$, is required to explain the particle
spectra at SPS energy, see Fig. \ref{spsmtspectra}.

In summary, we have formulated a statistical model of hadron resonance formation and
decay. Within the model, the density of hadronic states in mass is described
in terms of a universal Hagedorn-type temperature.
We have demonstrated that both the RHIC and SPS abundance data can be suitably described
in terms of resonance decays at the spectral temperature of $T_H \simeq 170$~MeV,
even when pursuing the extreme assumption of a single heavy resonance populating
the investigated rapidity region. To explain the data for particle spectra, we
needed to invoke additional collective motion beyond that generated in the hadronic interactions.

\bigskip

\begin{acknowledgments}
This work was supported by the U.S.\ National Science
Foundation under Grant PHY-0245009 and by the U.S.\ Department of Energy under
Grant DE-FG02-03ER41259.
\end{acknowledgments}


\begin{thebibliography}{99}
\bibitem{feed} As in the data, the results for $\overline B/B$ ratios are corrected
for feed down from weak decays, while the ratios involving particles with
$h^-$ and $\pi^-$ include weak decays from heavier hyperons.
\end{thebibliography}
\end{document}